\documentstyle[prb,aps,preprint,psfig]{revtex}
\tighten
\begin{document}
\draft 
\title{A Wannier function based {\em ab initio} 
Hartree-Fock approach extended to polymers: applications to the LiH chain
and {\em trans}-polyacetylene}
\author{Alok Shukla\cite{email}, Michael Dolg} 
\address{Max-Planck-Institut f\"ur
Physik komplexer Systeme,     N\"othnitzer Stra{\ss}e 38 
D-01187 Dresden, Germany}
\author{Hermann Stoll} \address{Institut f\"ur Theoretische Chemie,
Universit\"at Stuttgart, D-70550 Stuttgart, Germany}

\maketitle

\begin{abstract}
A recently proposed {\em ab initio} Hartree-Fock approach aimed
at directly obtaining the Wannier functions 
of a crystalline insulator is applied to polymers. The systems
considered are the LiH chain and {\em trans}-polyacetylene.
In addition to being the first application of our approach to one-dimensional
systems, this work also demonstrates its applicability 
to covalent systems. Both minimal as well as extended
basis sets were employed in the present study and excellent agreement
was obtained with the Bloch orbital based approaches. Cohesive
energies, optimized lattice parameters and the band structure
are presented. Localization characteristics of the Wannier functions are also
discussed.
\end{abstract}
\pacs{ }
\section{INTRODUCTION}
\label{intro}

Polymers represent a class of one-dimensional infinite crystalline systems 
where {\em ab initio} Hartree-Fock (HF) methods are well 
developed~\cite{hf-form,hf-karp,lih-karp,tpa-karp,hf-del,lih-del,tpa-kert,%
hf-suhai,hf-suhai2,hf-dovesi,lih-andre,hf-teramae,hf-teramae2,lihtpa-teramae}.
In addition, several groups have gone 
beyond the 
 HF level and have included the influence of electron correlations as
well~\cite{corr-ladik,corr-suhai0,corr-suhai,corr-suhai2,corr-lieg,%
corr-koenig,corr-sun0,corr-sun,corr-foerner,corr-foerner2,corr-myu}. Owing to
the reduced dimensionality, generally,
the computational effort involved in an {\em ab initio}
study of a polymer is considerably less than that in the case
of a corresponding three-dimensional (3D) crystal. 
It is possible, therefore, to perform high-quality 
{\em ab initio} calculations for polymers which are generally not 
yet feasible for 3D solids.

Barring a few exceptions involving studies using
Wannier-type orbitals~\cite{corr-suhai,corr-suhai2,corr-foerner2,poly-wan}, 
the method of choice to study the 
electronic structure of polymers is based on the use of Bloch 
orbitals.
Even where Wannier functions (WF) were used, they were 
obtained by {\em a posteriori} localization of 
Bloch orbitals. Recently, we have proposed an approach to the
electronic structure of periodic insulators which is formulated entirely
in terms of Wannier functions~\cite{shukla1,shukla2,albrecht,shukla3}, without
using Bloch orbitals in any of the intermediate steps. The theory
underlying this approach, which deals with the direct determination of the
Hartree-Fock Wannier orbitals of a crystalline insulator, is treated in 
detail in refs.~\cite{shukla1,shukla2}. The equivalence of the
Wannier-function-based approach to the Bloch-orbital-based approach at the
HF level was demonstrated  for quantities as diverse as total energy,
X-ray structure factors, Compton profiles, band structure, bulk modulus etc.\
of some 3D ionic 
insulators such as LiH~\cite{shukla1}, LiF~\cite{shukla2}, 
LiCl~\cite{shukla2}, NaCl~\cite{albrecht}, Li$_2$O~\cite{shukla3} and 
Na$_2$O~\cite{shukla3}.
However, owing to the highly localized nature of electronic states in ionic compounds,
they are naturally more amenable to a Wannier-function-based approach than
covalent systems. 
Therefore, by presenting HF calculations for {\em trans}-polyacetylene, the
aim of the present work is not only to demonstrate
the applicability of our approach to periodic systems of reduced 
dimensionality, but also the ease with which the present 
approach can be applied to study {\em covalent} systems. In addition to
polyacetylene, we also present calculations on a model ionic polymer, namely, 
an infinite LiH chain.
Since for ionic systems the long-range electrostatic contributions are very 
important, an accurate treatment of the Coulomb lattice sums becomes of crucial 
importance here; by comparing our results with those of other authors,
we can gauge the accuracy of the treatment of long-range Coulomb interaction in our
work. In all the calculations both minimal and extended basis sets were used.
For LiH, only total energies at the optimized lattice constants were 
computed. For polyacetylene, in addition, we present the 
detailed band structure and cohesive energy.
Our main motivation behind adopting a  Wannier function based approach is, of 
course, its possible use in an {\em ab initio} treatment of
electron correlation effects in infinite periodic systems. This aspect
of our work will be explored in the next paper in this series.
In addition, the Wannier functions also offer the possibility of
an {\em ab initio} determination of parameters involved in various
model Hamiltonians formulated in terms of localized orbitals such as
the H\"uckel model~\cite{hueckel}, the Hubbard model~\cite{hubbard} and the 
Pariser-Parr-Pople (PPP) model~\cite{ppp}.
We will also investigate these possibilities in a future publication.

The remainder of the paper is organized as follows. In Section \ref{theory}, 
we briefly sketch the theory with particular emphasis on the treatment of the
Coulomb lattice sums which differs from our Ewald-summation based approach
adopted for the 3D crystals. In Section \ref{results}, we present the results
of our calculations, while Section \ref{conclusion} contains our conclusions.
\section{THEORY}
\label{theory}
\subsection{Hartree-Fock Equations}
\label{hft}

For the sake of completeness, in this section we briefly review the underlying
theory in an intuitive manner. Rigorous derivations, along with 
details pertaining to the computer implementation, can be found in our
previous papers~\cite{shukla1,shukla2}. To solve the Hartree-Fock problem
of an infinite periodic system in the Wannier representation (as against
the traditional Bloch representation) we adopt a ``divide and conquer''
strategy. In this approach, we partition the infinite system into
a reference cell called the central cluster ($\cal{C}$), and its
environment (${\cal E}$) consisting of the rest of the infinite number of
unit cells. Thus, we can envision ${\cal C}$ as a cluster embedded in the field
of the rest of the infinite solid. Since the translational symmetry requires
that the orbitals localized in two different unit cells be identical
to each other (except for their location), it clearly suffices for us
to know the orbitals of the central cluster only, whereas the orbitals of all 
other cells can be generated from
them by simple translation operations. 
If we restrict the use of the Greek indices $\alpha$,
$\beta$ and $\gamma$ etc. to denote the (occupied) Wannier orbitals of the reference cell, and
accordingly choose the set $\{ |\alpha\rangle; \alpha =1,n_{c} \}$ to
represent the Wannier functions of the $2n_{c}$ electrons localized in
${\cal C}$, then the condition of translational symmetry can be expressed as
\begin{equation}
|\alpha({\bf R}_{i})\rangle = {\cal T} ({\bf R}_{i}) 
|\alpha({\bf 0})\rangle \mbox{,}
\label{eq-trsym}
\end{equation}
where $|\alpha({\bf 0})\rangle$ represents a Wannier orbital localized in
the reference unit cell assumed to be located at the origin while  $|\alpha({\bf R}_{i})\rangle$ is the corresponding orbital
of the $i$-th unit cell located at lattice vector ${\bf R}_{i}$, 
and the corresponding translation
is induced by the operator ${\cal T} ({\bf R}_{i})$. This immediately
suggests an iterative self-consistent-field (SCF) procedure. We can start
the calculations with a reasonable starting guess for the orbitals of
${\cal C}$, and consequently those of ${\cal E}$. These orbitals, in turn,
can be used to set up the embedded-cluster Hamiltonian for the electrons
of ${\cal C}$, which, upon diagonalization, leads to a new set of orbitals.
This procedure can be iterated until self-consistency
is achieved indicated by a converged value of the total energy per cell. 
Clearly, the above mentioned SCF procedure is applicable to any
independent-particle effective Hamiltonian such as the Kohn-Sham or the
Hartree-Fock Hamiltonian. However, in what follows, we will focus exclusively
on the {\em ab initio} restricted Hartree-Fock (RHF) implementation of the embedded-cluster
approach outlined above. In our previous work we showed that one
can obtain a set of RHF Wannier  functions of the $2n_{c}$ electrons 
localized in ${\cal C}$ by solving the equations~\cite{shukla1,shukla2} 
\begin{equation}
( T + U
 +   \sum_{\beta} (2 J_{\beta}-  K_{\beta})   
+\sum_{k \in{\cal N}} \sum_{\gamma} \lambda_{\gamma}^{k} 
|\gamma({\bf R}_{k})\rangle
\langle\gamma({\bf R}_{k})| ) |\alpha\rangle
 = \epsilon_{\alpha} |\alpha\rangle
\mbox{,}
\label{eq-hff1}         
\end{equation}  
where $T$ represents the kinetic-energy operator, $U$ represents
the interaction of the electrons of ${\cal C}$  with the nuclei
of the whole of the polymer while $J_{\beta}$, $K_{\beta}$  defined as 
\begin{equation}
\left.
 \begin{array}{lll}
 J_{\beta}|\alpha\rangle & = & \sum_{j} \langle\beta({\bf R}_{j})|\frac{1}{r_{12}}|
\beta({\bf R}_{j})\rangle|\alpha\rangle \\  
 K_{\beta}|\alpha\rangle & = & \sum_{j} \langle\beta({\bf R}_{j})|\frac{1}{r_{12}}|\alpha\rangle
|
\beta({\bf R}_{j})\rangle \\  
\end{array}
 \right\}  \mbox{,} \label{eq-jk} 
\end{equation}
respectively incorporate the Coulomb and exchange interactions of the 
electrons of ${\cal C}$ with those of the infinite system.
The first three terms of Eq.(\ref{eq-hff1}) constitute the canonical Hartree-Fock
operator, while the last term is a projection
operator which makes the orbitals localized in ${\cal C}$ orthogonal to those 
localized in the unit cells in the immediate neighborhood of ${\cal C}$
by means of infinitely high shift parameters $\lambda_{\gamma}^{k}$'s. These
neighborhood unit cells, whose origins are labeled by lattice vectors
${\bf R}_{k}$, are collectively referred to as  ${\cal N}$. The 
projection operators along with the shift
parameters play the role of a localizing potential in the Fock matrix, and 
once self-consistency has been achieved, the occupied eigenvectors of 
Eq.(\ref{eq-hff1})  are localized in ${\cal C}$, and are orthogonal to the 
orbitals of ${\cal N}$---thus making them Wannier 
functions~\cite{shukla1,shukla2}. As far as the
orthogonality of the orbitals of ${\cal C}$ to those contained in unit cells
beyond ${\cal N}$ is concerned, it should be automatic for systems with
a band gap once  ${\cal N}$ has been chosen to be large enough. In what
follows we shall specify the size of ${\cal N}$ by specifying the number
$N$ which implies the number of nearest neighbors that are included in
${\cal N}$. For example, $N=3$ shall imply that ${\cal N}$ contains up to
third-nearest neighbors of ${\cal C}$, and so on. The influence of the choice
of $N$ on the results of the calculations will also 
be studied in section~\ref{results}.
 
We have computer-implemented the formalism outlined above within a 
linear combination of atomic orbitals (LCAO) scheme, utilizing 
Gaussian-lobe-type basis functions~\cite{wannier}.  
We proceed by expanding the orbitals localized in the reference cell as~\cite{shukla2}
\begin{equation}
|\alpha\rangle =  \sum_{p} \sum_{{\bf R}_{j} \in {{\cal C} + \cal N} } 
C_{p({\bf R}_{j}),\alpha} |p({\bf R}_{j})\rangle  \: \mbox{,}
\label{eq-lcao}
\end{equation}
where ${\cal C}$ has been used to denote the reference cell, ${\bf R}_{j}$ 
represents the location of the  
$j$th unit cell (located in ${\cal C}$ or ${\cal N}$)  
and $|p({\bf R}_{j})\rangle$ represents
a lobe-type basis function centered in the $j$th unit cell. In order to account for the
orthogonalization tails of the reference cell Wannier orbitals, it is 
necessary to include the basis functions centered in ${\cal N}$ as well. 
The main aspect which
makes the problem of the infinite solid different from the problem
of a molecule that one usually encounters in quantum chemistry, is the
presence of infinite lattice sums in the terms  $U$, $J$ and $K$ of 
Eq.(\ref{eq-hff1}). Of these, the exchange interaction depicted by $K$
is fairly short-range for insulators, and converges rapidly. However,
the terms $U$ and $J$ involve long-range Coulomb interactions and are
individually divergent. Therefore, they need special consideration.
In our work on 3D insulators published 
earlier~\cite{shukla1,shukla2}, we resorted to the Ewald-summation technique
in order to evaluate these contributions. But, for the case of one-dimensional
systems considered here, we use a completely real-space
summation approach to be discussed in the next subsection.

To obtain the band structure we adopt the approach outlined in our previous
work~\cite{albrecht}. This essentially consists of first Fourier transforming 
the converged real-space Fock matrix (cf. Eq.(\ref{eq-hff1})) to gets its
k-space representation, and then rediagonalizing it to obtain the band energies
and eigenvectors. 
\subsection{Treatment of the Coulomb Series}
\label{coulser}
The matrix elements of electron-nucleus interaction that one needs to
construct the  LCAO version of Eq.(\ref{eq-hff1}) for the case of a polymer 
are~\cite{shukla2}
\begin{equation}
 U_{pq}({\bf t}_{pq})=-\sum_{j=-M}^{M} \sum_{A}^{\mbox{atoms}}
\langle p({\bf t}_{pq}) | \frac{Z_{A}}{|{\bf r} - {\bf R}_{j} -{\bf r}_A|} | q(0)
\rangle \; \mbox{,} 
\label{upq}
\end{equation}
where $|p({\bf t}_{pq}\rangle$ and $|q(0)\rangle$ denote two basis functions separated by an arbitrary vector
of the lattice ${\bf t}_{pq}$. ${\bf R}_{j}$
denotes the location of a unit cell, $Z_{A}$ represents the nuclear charge
of the $A$-th atom of the unit cell, ${\bf r}_A$ represents its fractional
coordinates, and the summation over  $A$ naturally runs over
all the atoms in the unit cell. Of course, for an infinite polymer 
$M \rightarrow \infty$. Similarly, to describe the Coulombic part of
the electron-electron repulsion, we need matrix elements of the form
\begin{equation}
J_{pq;rs}({\bf t}_{pq},{\bf t}_{rs}) = \sum_{j=-M}^{M} 
\langle p({\bf t}_{pq}) \:r({\bf t}_{rs}+{\bf R}_{j}) |\frac{1}{|{\bf r}_1 
- {\bf r}_2| }
|q(0)  \: s({\bf R}_{j}) \rangle
\; \mbox{,} \label{jpqrs1}
\end{equation}
which, by means of a coordinate transformation, can be brought into a form
very similar to that of Eq.(\ref{upq})
\begin{equation}
J_{pq;rs}({\bf t}_{pq},{\bf t}_{rs}) = \sum_{j=-M}^{M} 
\langle p({\bf t}_{pq}) \:r({\bf t}_{rs}) |\frac{1}{|{\bf r}_1 
- {\bf r}_2 - {\bf R}_{j}| }
|q(0)  \: s(0) \rangle
\; \mbox{.} \label{jpqrs2}
\end{equation}
Although the individually infinite series involved in Eqs. (\ref{upq}) and
(\ref{jpqrs2}) are divergent, they can be forced to converge by means
of the Ewald-summation method~\cite{ewald,ew-stoll}. However, if one
uses one and the same, sufficiently large value of $M$ to directly 
evaluate the 
matrix elements of Eqs. (\ref{upq}) and (\ref{jpqrs2}), the divergences
inherent in the two terms will cancel each other owing to the opposite
signs when combined together to form the corresponding Fock matrix element.
The total energy per unit cell will also be convergent if one uses the
same value of $M$ to evaluate the contribution of the nucleus-nucleus
interaction energy as well. Besides the finite lattice sums over
the unit cell index $j$ in the equations above, we have not included
any other long-range corrections such as ones based upon 
multipole expansions~\cite{lih-del}. The real-space approach outlined above is
similar in spirit to the one used by Dovesi in his Bloch orbital based
study of polyacetylene~\cite{hf-dovesi}. Most of the other
authors also adopt the real-space based summation of the Coulomb series
to perform {\em ab initio} studies on 
polymers~\cite{hf-form,hf-karp,lih-karp,tpa-karp,hf-del,lih-del,tpa-kert,%
hf-suhai,hf-suhai2,hf-dovesi,lih-andre,hf-teramae,hf-teramae2,lihtpa-teramae}
 However, these schemes differ in
various details related to the cutoff used in the truncation of the series.
The convergence properties of the total energy per unit cell and, to some 
extent its final value, are 
frequently dependent on the scheme adopted.
For an excellent account of different cutoff schemes in practice, and their
convergence properties, we refer the reader to a recent article by
Teramae~\cite{lihtpa-teramae}.

In the present scheme we calculate only the set of integrals indicated by 
Eqs. (\ref{upq}) and (\ref{jpqrs2}) and generate all the integrals needed
from this set by using translational invariance. However, strictly speaking,
the translationally invariant form of these equations 
is valid only in the limit $M \rightarrow\infty$. Since all the calculations
presented in this work are restricted to finite values of $M$, the use
of translational invariance embodied in Eqs. (\ref{upq}) and (\ref{jpqrs2}) 
is an approximation. Therefore, it is important to  study carefully the
convergence of the total energy per unit cell as a function of $M$ and we
will present our findings in Sec.~\ref{results}. 
\section{CALCULATIONS AND RESULTS}
\label{results}
In this section we present the results of calculations performed on both
the model polymer LiH and the ''real'' polymer {\em trans}-polyacetylene. To
check the accuracy of our approach, we also performed the same
calculations with the CRYSTAL program~\cite{crystalprog} and will
present those results as well. Since our approach does not include
the long-range corrections to the Coulomb interaction, while the
CRYSTAL program does include them via the Ewald summation, we believe
that this comparison is quite instructive. Wherever possible, we will
also compare our results to those of other authors.
\subsection{LiH}
Perhaps because of its simplicity, the LiH chain has been studied
by several authors prior to this work~\cite{lih-karp,lih-del,lih-andre,%
lihtpa-teramae}. The reason behind our study of this system is twofold. 
Firstly, being an ionic polymer,
the long-range Coulomb interactions are very important for the LiH chain. 
Since our approach does not rely on an infinite sum of this effect,
comparison of our results with those of the CRYSTAL 
program~\cite{crystalprog} will help us to judge the quality of our treatment of 
the Coulomb series. 
Secondly, as mentioned previously, our
program uses lobe functions to approximate the $p$ and higher angular momentum
cartesian basis functions. Since most of the other authors use
true cartesian basis functions, comparison between our results and those of 
other programs such as CRYSTAL~\cite{crystalprog} can only be approximate 
when such
basis functions are involved. However, the LiH chain can be described reasonably
well using only $s$-type basis functions, a case for which the lobe- and the
cartesian-type Gaussian basis functions are trivially equivalent. Therefore,
a comparison of our results for the LiH chain involving only $s$-type 
Gaussian basis functions with those of other authors,
will be a further test of the correctness of our approach.

Karpfen~\cite{lih-karp} and Delhalle et al. ~\cite{lih-del} concluded
that for an infinite LiH chain, the equilibrium geometry corresponds
to the case where Li and H atoms are equidistant from each other. We also
adopted a similar geometry, with the reference cell having
H at $(0,0,0)$ and Li at $(a/2,0,0)$, where $a$ is the lattice constant
of the chain. The 
chain was assumed to be oriented along the $x$ axis.

To study the LiH chain with a (sub)-minimal basis set, we adopted the
STO-4G basis set optimized by Dovesi et al.~\cite{lih-dovesi} for their study
of the bulk LiH. Thus, there are two basis functions per unit cell, 
with one basis function each on Li and H sites. With this basis set
we obtained an equilibrium lattice constant of 6.653 atomic units. The results
of our calculations at the equilibrium lattice constant, and its comparison 
with those of the CRYSTAL~\cite{crystalprog} program, are presented in the 
table \ref{tab-enlih1}. To the best of our knowledge, the STO-4G basis has 
not been used by any other author to study the LiH chain, so that for this
case our comparison is restricted only to the CRYSTAL~\cite{crystalprog} 
results.
For the extended basis set calculations we used
the contraction coefficients and the exponents reported by Huzinaga,
both for Li~\cite{huz-li} and H~\cite{huz-h}. The Li basis set was of the
type $(8s)/[5s]$ while the H basis set was of $(4s)/[3s]$ type with, in total, 
eight basis functions per unit cell. This basis set 
was also used by Delhalle et al.~\cite{lih-del} in their study of the LiH infinite
chain, employing a multipole-expansion-oriented approach for the
Coulomb series. They obtained an equilibrium lattice constant of 6.478 a.u.,
 which is the value that we have also used to perform
our computations presented in table \ref{tab-enlih2}. In the same table, our
results are compared with those of Delhalle et al.~\cite{lih-del} and
those obtained using the CRYSTAL program. In every calculation
involving either our program or CRYSTAL~\cite{crystalprog}, all the
one- and two-electron integrals whose absolute value was below $10^{-7}$ a.u.
were discarded. 

From tables \ref{tab-enlih1} and \ref{tab-enlih2} one can easily understand
the convergence pattern of our results as far as its dependence on the size of
the orthogonality region ${\cal N}$, and the number of neighbors in the
Coulomb series $M$, is concerned. A quick glance at both the tables reveals 
that it is not
sufficient to orthogonalize the Wannier orbitals of the reference cell
${\cal C}$ only to those in its nearest neighbor cells ($N=1$). As is clear,
the lack of sufficient orthogonality for those cases leads to energies
lower than the true energies. However,
if we orthogonalize the Wannier orbitals of ${\cal C}$ to, at least,
those in the second-nearest-neighbor cells ($N=2$), we attain convergence in
total energy per unit cell. This fact is obvious by noticing that the
energies obtained with the orthogonality requirement restricted to the
second-nearest neighbors ($N=2$) agree at the micro-Hartree level with those 
obtained when the orthogonality requirement was extended to the third- ($N=3$)
 and the fourth-nearest neighbors ($N=4$), respectively. This rather fast 
convergence with respect to $N$ can be understood intuitively if one consisders
the fact that the LiH chain is a large-band-gap insulator. This, in turn, points
to the well-localized character of the valence electrons residing
predominantly on the H$^-$ Wannier functions. With well-localized valence
electrons, one should not expect them to have sizeable overlaps with the
electrons localized in the far-away unit cells.

Now we examine the convergence of the results with respect to the number of
neighbors $M$ included in the Coulomb series. For the reasons mentioned above,
we will only consider those of our results which correspond to $N=2$ or 
higher. Even a cursory inspection of
tables \ref{tab-enlih1} and \ref{tab-enlih2} reveals that, as expected, this 
convergence, is much slower as compared to the one with respect to $N$. This 
can also be intuitively understood as a consequence of the long-range
character of the Coulomb interactions in an ionic system like the LiH chain.
Indeed, we find for the case of the minimal basis set that our results are
1 microHartree off the CRYSTAL results. This small disagreement could
also be due to some numerical error in either of the codes. For the
case of extended basis set we have exact microHartree-level agreement  
with the results of Delhalle et al.~\cite{lih-del} and 
CRYSTAL~\cite{crystalprog}, once well-above 200 nearest-neighbors have been 
included in the Coulomb series. However, for evaluating energy differences in quantum-chemical
calculations, it is often sufficient to have results accurate up to 1 milliHartree.
As is clear from both the tables, this level of accuracy is achieved
with about 40 neighbors included in the Coulomb series. 
Thus the fact remains that in absolute terms the Coulomb series converges
quite slowly; however, for the purpose of a calculation with reasonable
accuracy, the computational effort involved in a direct scheme as outlined
in Sec. ~\ref{coulser} is not too prohibitive. 
\subsection{{\em Trans}-polyacetylene}
\label{sec-tpa}
The isomer {\em trans}-polyacetylene(t-PA) has been the subject of numerous 
studies, both at the Hartree-Fock~\cite{tpa-karp,tpa-kert,hf-suhai,hf-suhai2,%
hf-dovesi,hf-teramae2,lihtpa-teramae} and at correlated 
levels~\cite{hf-suhai2,corr-suhai0,corr-suhai2,corr-lieg,corr-koenig,%
corr-sun0,corr-foerner}.
It has an alternant structure as shown in Fig. \ref{fig-tpa}, with the 
length of the double bond ($r_{2}$)
being shorter than that of the single bond ($r_{1}$). 
The difference in 
the corresponding bond lengths $\Delta r= r_1 - r_2$ is called the  
{\em bond alternation}. 
If the two bond lengths were equal, i.e., a zero bond alternation, 
the unit cell of t-PA will consist of a single CH unit giving it 
a metallic character with a half-filled $\pi$ band. However, 
in reality, because of nonzero bond alternation, t-PA has a dimerized unit
cell consisting of a C$_2$H$_2$ unit which naturally leads to insulating
behavior.   The dimerization is widely believed to be a consequence of Peierls 
distortion which follows from the coupling of the phonons to electrons on the 
Fermi surface~\cite{peierls}. The phenomenon
of nonexistence of one-dimensional metals due to 
Peierls distortion---sometimes also referred to as Peierls dimerization---has
come to be known as the Peierls theorem.

In this work we have used the lobe representations of both the minimal 
STO-3G~\cite{sto-3g} basis set as well as the extended 
6-31G basis set, to optimize the
geometry and to obtain the cohesive energies at the Hartree-Fock level.
We have used the 6-31G basis set in two versions. Since the use of d-type
functions in a lobe representation is computationally very expensive,
we have dropped polarization functions on the carbon atoms during our study
of the convergence pattern of the Coulomb series with the extended basis
set, although polarization functions on the hydrogen atoms were retained. 
The polarization function on hydrogen
consisted of a single p-type exponent of 0.75 a.u.
From now on, we refer to this restricted form of the 6-31G basis set with a [3s,2p] basis
set on carbon and a [2s,1p] set on hydrogen as the 6-31G-1 basis set. For the
geometry optimization and band structure calculations, we augmented the carbon
basis by one d-type exponent of 0.55 a.u. and refer to the basis set by its
conventional name of 6-31G**.  
For the sake of comparison, we 
also performed the same set of calculations with the CRYSTAL program. 
As in the case of LiH, in both our and CRYSTAL calculations all the one- and
two-electron integrals with magnitude less than $1.0\times 10^{-7}$ a.u. were
neglected.
 In the present calculation, the C-H bond length
was assumed to be fixed at the experimental value of 1.09 $\AA$ and the
reference unit cell was assumed to be a dimerized primitive cell consisting
of a C$_2$H$_2$ unit, also shown in Fig. \ref{fig-tpa}. For optimizing the
geometry, the bond lengths $r_1$,
$r_2$, and the bond angle $\alpha$ between the two C-C  
bonds were allowed to vary.

To study t-PA at the HF level the STO-3G basis sets have been used earlier by 
Kert\'{e}sz et al.~\cite{tpa-kert}, Suhai~\cite{hf-suhai,hf-suhai2}, 
Karpfen et al.~\cite{tpa-karp}, Dovesi~\cite{hf-dovesi} and recently
by Teramae~\cite{lihtpa-teramae}.  Teramae~\cite{lihtpa-teramae} and 
Suhai~\cite{hf-suhai2} in addition to other lattice 
parameters also optimized the
C-H bond length which was found to be different from the value
1.09 $\AA$ used in the present work (as well as by other authors mentioned 
above).
Therefore, we cannot directly compare our results to those of Teramae
and Suhai. Of the
other authors, only  Karpfen
et al.~\cite{tpa-karp} and Dovesi~\cite{hf-dovesi} performed the
geometry optimization.  The optimized values for $r_{1}$, $r_{2}$ and the 
bond angle $\alpha$ were obtained to be respectively
1.477$\AA$, 1.327$\AA$ and 124.2$^{\circ}$ by Karpfen et al.~\cite{tpa-karp}
and 1.486$\AA$, 1.329 $\AA$ and 124.4$^{\circ}$ by Dovesi~\cite{hf-dovesi}.
The optimized values of 1.489$\AA$, 1.326$\AA$ and 124.1$^{\circ}$ obtained 
by us in the present work clearly are in good agreement with the previous 
results. 

With the extended 6-31G-1 basis set, the optimized values of $r_1$, $r_2$ and
$\alpha$ obtained with our approach were 1.452$\AA$, 1.340 $\AA$ and 
124.4$^{\circ}$. When we performed the geometry optimization with the
same basis set using the CRYSTAL program we obtained 1.458$\AA$, 1.336 $\AA$ 
and 124.5$^{\circ}$ for these quantities. When we used the 6-31G** basis
set for the same task, the optimized
values with our program were 1.457$\AA$, 1.336 $\AA$ and 124.2$^{\circ}$, and with the CRYSTAL
code we determined them to be 1.464$\AA$, 1.333 $\AA$ and 124.2$^{\circ}$.
Clearly, for both types of extended basis sets, i.e., with and without 
polarization functions on the carbon atoms, there is excellent agreement 
between our optimized geometries and those obtained using the CRYSTAL 
program.

The convergence pattern of the total energy per unit
cell at the optimized geometries mentioned above, as a function of 
the parameters $M$ and $N$ is
displayed in table \ref{tab-entpa1} for the STO-3G set and in table 
\ref{tab-entpa2} for the 6-31G-1 basis set. Contrary to 
the case of the LiH chain,
we were not able to achieve convergence if the orthogonality region
of the Wannier functions was smaller than the third-nearest neighbors ($N=3$).
This observation can be understood on the physical grounds that the Wannier
functions of a covalent system like t-PA are much
more delocalized as compared to an ionic system such as the LiH chain. 
Therefore,
 their orthogonalization tails extend much more into the neighborhood
than those of the Wannier functions of LiH. Although a micro-Hartree
level convergence in total energy is achieved only after including 
at least six nearest-neighbor cells in the orthogonality region, the
difference in total energy between the $N=3$ and $N=6$ cases is only $\approx$
24 micro Hartrees. Thus the convergence in the total
energy with respect to $N$ is quite rapid.

Similarly, for t-PA the convergence of the total energy per unit cell 
with respect to the number of nearest neighbors ($M$) included in
the Coulomb series turns out to be slower than for the LiH chain. As is clear from 
tables \ref{tab-entpa1} and \ref{tab-entpa2}, to achieve
a milliHartree convergence in the results, one needs to have at least
$M=75$, while the microHartree convergence is not achieved even after 
including 500 nearest neighboring cells. With the extended 6-31G-1 basis
set we did not achieve any convergence for the cases with $M=10$ and $M=20$.
This behavior is to be contrasted
with the case of the LiH chain (tables \ref{tab-enlih1} and \ref{tab-enlih2})
where $M=20$ sufficed for a milliHartree level of convergence and about
$M=200$ brought the results to within 1 microHartree of the converged
results.  Moreover, in most of the prevalent real-space
based approaches to the Coulomb series one observes much faster convergence
of the total energy, with reasonable results obtainable even for 
$M=3$ case~\cite{lihtpa-teramae}. Comparatively speaking, the slow convergence
of the Coulomb series observed by us appears contradictory. However, 
the reason behind this can be readily understood if one recognizes the 
primitive nature of the truncation criteria embodied in 
Eqs. (\ref{upq}) and (\ref{jpqrs2}). This cutoff scheme clearly pays little
regard to the charge balance in the unit cell. In addition, it 
uses translational invariance when, in reality, it is strictly valid
only in the limit $M \rightarrow \infty$. Since charge distributions for
t-PA are much more delocalized than LiH, any charge imbalance should lead
to slower convergence in the former case. This is consistent with our
observations. In such a case, would also expect the error due to charge
imbalace to diminish with increasing value of $M$, again consistent with
our observations.
However, one could accelerate the convergence of the Coulomb series in a 
computationally inexpensive manner either by adopting an Ewald-summation based 
approach~\cite{ew-stoll} or by using a multipole expansion based 
approach~\cite{lih-del}. 
Anyway, for crystalline systems typically  results accurate up to
milliHartree  level are sufficient, and that level of convergence is achieved
by using $M=75$ which is computationally not too expensive. 
The comparison of our total energy per unit cell with that obtained using 
the CRYSTAL program employing the identical geometry is excellent to within
a few fractions of a milliHartree. We observed the same level of 
(dis)agreement with the CRYSTAL results in our previous studies on
3D solids~\cite{shukla1,shukla2,albrecht,shukla3}, where we had used the
Ewald summation approach to treat the Coulomb series. This gives
us confidence that the small disagreements in the total energy per unit
cell with respect to the CRYSTAL results are largely due to our use of lobe 
functions to approximate the cartesian-type
Gaussian basis functions used in the CRYSTAL program. Therefore, we believe,
that the treatment of the Coulomb series outlined in the present work, although
slowly convergent, is conceptually on sound foundations.

We also evaluated the band structure of t-PA, at the most recently
reported experimental 
geometry~\cite{tpa-exp} with $r_1=1.45$ $\AA$, $r_2=1.36$ $\AA$ and
the lattice constant of 2.455 $\AA$ which corresponds to a bond angle
$\alpha=121.7^{\circ}$, using the 6-31G** basis set.  For these
calculations the choice of orthogonality parameter was $N=3$  and 
the Coulomb-series parameter was $M=100$. The four highest occupied
bands, along with the five lowest conduction bands are plotted
in Fig. \ref{fig-band}. The same figure also plots the corresponding
bands obtained using the CRYSTAL program employing the same basis set
and geometry. The absolute values of the band energies naturally 
differed somewhat owing to the different treatment of the Coulomb series in the
two approaches. Therefore we shifted all the CRYSTAL band energies so that 
the tops of the valence bands obtained from the two approaches coincided.
Clearly, the band structures obtained from the two approaches are in 
excellent agreement for the occupied bands and for the lowest three conduction
bands. The value of the direct band gap (at $k=\frac{\pi}{a}$ point) 
0.2356 a.u. (6.41 eV)  obtained with our approach is in good agreement with
the corresponding CRYSTAL value of 0.2339 a.u. (6.37 eV).  
For the fourth and the fifth conduction bands we see some small deviations. 
For the higher conduction bands not plotted in Fig. \ref{fig-band}, 
the deviations
are even more significant. However, this behavior is to be expected when
one uses lobe functions because, even for molecular systems, unoccupied
energy levels generally differ significantly from each other when the
same calculation is performed with lobe- and the cartesian-type 
functions. We saw a similar trend in our earlier work on the band structure
of the NaCl crystal~\cite{albrecht}. The experimental value of the direct 
gap is widely believed to be $\approx$ 2 eV~\cite{tpa-band}. Therefore, as is
generally the case with HF bands, the band gap of t-PA 
is overestimated by a large amount, pointing to the importance of
the electron correlation effects. The influence of electron correlations
on the band structure of t-PA has been studied by Suhai~\cite{corr-suhai0},
Liegener~\cite{corr-lieg} and by Sun et al.~\cite{corr-sun0} within Bloch
orbital based approaches. F\"orner et al.~\cite{corr-foerner2} have recently 
included the electron-correlation effects in the band structure using a 
Wannier-function-based coupled-cluster approach. All the prior studies
indicate that once the electron correlations are accounted for, one observes
a dramatic reduction in the band gap.

Our results for ground-state properties with the STO-3G basis set are summarized in table
\ref{tab-sto3g}. This table also presents results of other authors who performed
calculations using the same basis set. Noteworthy entries in the table 
are the results of recent calculations by Teramae~\cite{lihtpa-teramae} which 
were performed using different cutoff schemes for the treatment of the 
Coulomb series. The details of these cutoff schemes can be obtained in the
above-mentioned paper or in the original papers
cited therein. The differences in the results with the same basis set
but with different cutoff schemes
clearly testify to the fact that the treatment of the Coulomb series
is a delicate matter which deserves utmost caution. Our own view is that
unambiguous results will only be obtained when the Coulomb series is
treated in the Ewald limit as is done, e.g. in the CRYSTAL 
program~\cite{crystalprog}, or by saturating the Coulomb series to 
a very large number of unit cells which can be done inexpensively, e.g,
by using the multipole expansion techniques of Delhalle et al.~\cite{lih-del}.
In our opinion, these schemes should be treated as standard, and the rest
of the prevalent schemes should be judged against them.

Our final results obtained with the extended 6-31G** basis set are 
presented in table \ref{tab-tpaf} which also  compares them to the
calculations performed by us---employing the same basis set---with
the CRYSTAL program. The table also presents the results of 
Suhai~\cite{hf-suhai2} and of Yu et al.~\cite{corr-myu} which were all
performed with basis sets of similar quality as those used
by us. To evaluate the cohesive energies corresponding to our calculation,
we used Hartree-Fock reference energies for carbon and hydrogen of
-37.677838 a.u. and -0.498233 a.u., respectively. These energies were obtained
by performing atomic HF calculations employing the same 6-31G basis set as used in
the polymer calculations. It is apparent from the table that the results for cohesive energies
obtained by different authors, employing different methods and basis sets,
are in good aggreement. To the best of our knowledge, no experimental data
on the cohesive energy of t-PA are available.
However, it is well-known that the
HF method systematically underestimates the cohesive energy and, therefore,
one expects electron correlations to contribute significantly to the
true cohesive energy of t-PA.
The experimental geometry of t-PA is available from at least
three papers~\cite{tpa-exp,tpa-exp2,tpa-exp3} which disagree from
each other somewhat. However, we will use the most recent results of
Kahlert et al.~\cite{tpa-exp} as the reference.
Compared to experiment the HF calculations appear to 
overestimate the single-bond length $r_1$ and bond alternation $\Delta r$
by about 0.01 $\AA$ and 0.05 $\AA$ respectively, while the double-bond 
length $r_2$ is underestimated by at least 0.02 $\AA$. The bond-angle, which
is a measure of the lattice constant, is also overestimated at the HF level.
Therefore, the most significant deviation at the HF level is in the bond
alternation. Since the Peierls theorem, which predicts a nonzero bond alternation,
is an exact result only in the absence of electron correlations, 
it is of theoretical interest to study the influence of electron
correlations on the phenomenon of Peierls dimerization. 
The fact that the inclusion of electron correlations improves the agreement
with the experiment on all the geometry parameters including bond
alternation has been confirmed by K\"{o}nig et al.~\cite{corr-koenig} using
a ``local-ansatz'' based approach, by Suhai~\cite{corr-suhai} using
a Bloch-orbital-based MBPT approach, and by Yu et al.~\cite{corr-myu} using
an incremental scheme based local-correlation approach~\cite{increment}
applied to finite clusters simulating t-PA.

Finally a pictorial view of the Wannier function corresponding to the 
$\pi$ bond of the unit cell, evaluated at the experimental geometry, 
is provided in Fig. \ref{fig-pi}. The figure corresponds to the contour
plot of the charge density associated with the corresponding Wannier
function in the $xy$ plane with $z=0.25$ atomic units. From the contour 
plots the localized nature of the $\pi$ electrons, as well as their 
participation in a covalent bond between the two carbon
atoms of the unit cell, is obvious.

\section{CONCLUSIONS AND FUTURE DIRECTIONS}
\label{conclusion}
 In conclusion, an {\em ab initio} Wannier-function-based Hartree-Fock 
approach developed originally to treat infinite 3D crystalline systems 
has been extended to deal with polymers. The main difference as compared to
the case of 3D systems has been an entirely real-space based treatment of
the Coulomb series which has been demonstrated to be applicable both to
ionic and covalent systems. We observed slow convergence of the
Coulomb series with respect to the lattice sums, but this problem
can be rectified in the future by adopting either an Ewald-summation-based,
or a multipole-expansion-based approach to the Coulomb series. 
  
The main focus of this work was, of course, a detailed Hartree-Fock
study of {\em trans}-polyacetylene which involved the use of an extended
basis set including polarization-type functions. Various quantities such
as the total energy per unit cell, the cohesive energy, optimized geometry
parameters and the band structure were found to be in excellent agreement
with those found from equivalent calculations performed using the 
Bloch-orbital-based approach. In this manner we have demonstrated the
applicability of our approach to covalent systems where Wannier functions
are less well localized as compared to the ionic systems studied earlier by us.
One possible use of the present Wannier function based approach can be in
the theoretical determination of various parameters involved in
model Hamiltonians such as the H\"uckel Hamiltonian, the PPP and 
Hubbard models. For the particular case of $\pi$-electron systems such
as {\em trans}-polyacetylene for the description of which the PPP Hamiltonian
is frequently used, one can, after some numerical work, obtain a 
Hartree-Fock level estimate of the parameters involved. Such an estimate
can subsequently be refined by performing renormalization group procedures. We
will pursue this line of research in a separate publication.
The Wannier-function based approach  can also
be used to obtain insights into the various possible mechanisms, such
as soliton formation~\cite{su}, supposed to be behind the 
Peierls distortion of {\em trans}-polyacetylene. This can be done by
introducing the corresponding structural defect in a finite region 
around the reference cell, keeping the rest of the polymer frozen at the
level of the Hartree-Fock solution of the perfect polymer.

The discrepancy between our Hartree-Fock results for {\em trans}-polyacetylene
and the experimental ones was found to be most noteworthy for the bond
alternation and the band structure. These differences point to
the importance of electron-correlation effects. In a future
publication,  we will include these within
a local-correlation approach to study their effect on ground- and 
excited-state properties.
This way, it will be possible, in particular,
to study the influence of electron correlations on the Peierls
dimerization within an entirely real-space formalism in an {\em ab initio}
manner, which so far was usually restricted to model 
Hamiltonians~\cite{dixit}.

\begin{figure}
\caption{Structure of {\em trans}-polyacetylene as considered in the present
work. Bonds included in the reference cell ${\cal C}$ in the calculations
are enclosed in the dashed box.}
\label{fig-tpa} 
\end{figure}
\begin{figure}
\caption{Band structure of t-PA obtained using our approach (solid lines) 
compared to that obtained using the CRYSTAL program (dashed lines). The experimental
geometry~\protect\cite{tpa-exp} and a 6-31G** basis set was used in both cases. 
Values of $k$ (horizontal axis) are expressed in units of $\frac{2\pi}{a}$. 
The two sets of bands are essentially identical except for the 
top two conduction bands which are somewhat different.}
\label{fig-band} 
\end{figure} 
\begin{figure}
\caption{Contour plots of the charge density of the ${\pi}$-type 
valence Wannier function of the reference cell. Contours are plotted
in the $xy$ plane with $z=0.25$ a.u. ($x$ is the axis of the polymer). The
magnitude of the contours is on
a natural logarithmic scale. The two carbon atoms of the 
unit cell are located at the positions $(-1.11,0.64,0.0)$ a.u. and
$(1.11,-0.64,0.0)$ a.u. respectively.
Clearly the dominant contours are surrounding
the two carbon atoms of the reference cell indicating a covalent bond
between them. Weaker contours due to the orthogonalization tails of the
Wannier function extend up to nearest-neighbor carbon atoms and beyond.
The rapidly decaying strength of the contours testifies to the localized nature
of the Wannier function.}
\label{fig-pi} 
\end{figure}
\begin{table}  
 \protect\caption{Total energies per unit cell obtained in the present work, 
as a function of
the number of nearest-neighbor unit cells included in the Coulomb series, 
($M$), and those included in the orthogonality region ${\cal N}$ ($N$).  
For the sake of comparison, results of equivalent calculations performed with 
the CRYSTAL~\protect\cite{crystalprog} program, are 
also reported.
The STO-4G minimal basis
set of Dovesi et al.~\protect\cite{lih-dovesi} was used in all the calculations. All the results are
in atomic units, and refer to the optimized lattice constant of 6.653 a.u., with equidistant
Li and H atoms. }
 \protect\begin{center}  
  \begin{tabular}{cccccl} \hline
\multicolumn{5}{c}{This Work} &     CRYSTAL~\cite{crystalprog} \\ \cline{1-5}  
M  & \multicolumn{4}{c}{N}    &  \\ \cline{2-5}
   & 1 & 2 & 3 & 4 &   \\ \hline
10 & -7.997974 & -7.997898 & -7.997898 & -7.997898 &   \\
20 & -7.998249 & -7.998173 & -7.998173 & -7.998173 &  \\
30 & -7.998302 & -7.998227 & -7.998227 & -7.998227 &  \\
40 & -7.998322 & -7.998246 & -7.998246 & -7.998246 &   \\
50 & -7.998330 & -7.998255 & -7.998255 & -7.998255 &  \\
100 &-7.998343 & -7.998267 & -7.998267 & -7.998267 & \\
200 &-7.998346 & -7.998270 & -7.998270 & -7.998270 & \\ 
500 &-7.998346 & -7.998271 & -7.998271 & -7.998271 & \\ 
1000 
    &-7.998347 & -7.998271 & -7.998271 & -7.998271 & -7.998272 \\
\hline
   \end{tabular}                      
   \end{center}  
  \label{tab-enlih1}    
\end{table}  
\begin{table}  
 \protect\caption{Total energies per unit cell obtained in the present work, 
as a function of
the number of nearest-neighbor unit cells included in the Coulomb series, 
($M$), and those included in the orthogonality region ${\cal N}$ ($N$).  
For the sake of comparison, results of other authors are also reported.
Extended Huzinaga basis sets~\protect\cite{huz-li,huz-h} were used for Li and 
H in all the calculations. An optimized lattice constant of 6.478 a.u. was used 
along with the equidistant Li and H nuclei. 
All the results are in atomic units. }
 \protect\begin{center}  
  \begin{tabular}{cccccl} \hline
\multicolumn{5}{c}{This Work} &     Other Works  \\ \cline{1-5}  
M  & \multicolumn{4}{c}{N}    &  \\ \cline{2-5}
   & 1 & 2 & 3 & 4 &   \\ \hline
10 & -8.035526 & -8.035447 & -8.035447 & -8.035447 &   \\
20 & -8.035779 & -8.035701 & -8.035701 & -8.035701 &  \\
30 & -8.035829 & -8.035750 & -8.035750 & -8.035750 &  \\
40 & -8.035846 & -8.035768 & -8.035768 & -8.035768 &   \\
50 & -8.035855 & -8.035776 & -8.035776 & -8.035776 &   \\
100 &-8.035866 & -8.035788 & -8.035788 & -8.035788 & \\
200 &-8.035869 & -8.035790 & -8.035790 & -8.035790 & \\ 
500 &-8.035869 & -8.035791 & -8.035791 & -8.035791 & \\ 
1000 
    &-8.035869 & -8.035791 & -8.035791 &  -8.035791 & -8.035791$^{a,b}$ \\
\hline
   \end{tabular}                      
   \end{center}  
  \label{tab-enlih2}    
$^a$ ref.~\cite{lih-del} \\
$^b$ obtained using CRYSTAL program~\cite{crystalprog}.
\end{table}  
\begin{table}  
 \protect\caption{Total energies per unit cell for t-PA obtained in the present work, 
as a function of
the number of nearest-neighbor unit cells included in the Coulomb series, 
($M$), and those included in the orthogonality region ${\cal N}$ ($N$).  
For the sake of comparison, we also present results obtained with
the CRYSTAL program~\protect\cite{crystalprog}. In both the CRYSTAL and our
calculations, the STO-3G basis set along with the optimized geometry reported in 
Sec.\protect\ref{sec-tpa} were used.
All the results are in atomic units.}
 \protect\begin{center}  
  \begin{tabular}{ccccccl} \hline
\multicolumn{6}{c}{This Work} &     CRYSTAL  \\ \cline{1-6}  
M  & \multicolumn{5}{c}{N}    &  \\ \cline{2-6}
   & 3 & 4 & 5 & 6 & 7 &  \\ \hline
10 &-75.931783 &-75.931802 &-75.931807 &-75.931808 &-75.931808 & \\
20 &-75.943470 &-75.943489 &-75.943493 &-75.943494 &-75.943494 & \\ 
30 &-75.945851 &-75.945870 &-75.945874 &-75.945875 &-75.945876 & \\
40 &-75.946709 &-75.946728 &-75.946732 &-75.946733 &-75.946733 &   \\
50 &-75.947112 &-75.947131 &-75.947135 &-75.947136 &-75.947136 & \\
75 &-75.947514 &-75.947533 &-75.947537 &-75.947538 &-75.947538 & \\ 
100&-75.947656 &-75.947675 &-75.947680 &-75.947681 &-75.947681 & \\ 
200&-75.947794 &-75.947813 &-75.947818 &-75.947819 &-75.947819 & \\
300&-75.947820 &-75.947839 &-75.947843 &-75.947844 &-75.947844 & \\
400&-75.947829 &-75.947848 &-75.947852 &-75.947853 &-75.947853 & \\ 
500&-75.947833 &-75.947852 &-75.947856 &-75.947857 &-75.947858 & -75.947597 \\
\hline
   \end{tabular}                      
   \end{center}  
  \label{tab-entpa1}    
\end{table}  
\begin{table}  
 \protect\caption{Total energies per unit cell for t-PA obtained in the present work, 
as a function of
the number of nearest-neighbor unit cells included in the Coulomb series, 
($M$), and those included in the orthogonality region ${\cal N}$ ($N$).  
For the sake of comparison, we also present results obtained with
the CRYSTAL program~\protect\cite{crystalprog}. In both the CRYSTAL and our
calculations, the 6-31G-1$^a$ basis set along with the optimized geometry reported in 
Sec.\protect\ref{sec-tpa} were used.
All the results are in atomic units.}
 \protect\begin{center}  
  \begin{tabular}{ccccccl} \hline
\multicolumn{6}{c}{This Work} &     CRYSTAL  \\ \cline{1-6}  
M  & \multicolumn{5}{c}{N}    &  \\ \cline{2-6}
   & 3 & 4 & 5 & 6 & 7 &  \\ \hline
30 &-76.865184 &-76.865198 &-76.865204 &-76.865207 &-76.865207 & \\
40 &-76.865813 &-76.865826 &-76.865832 &-76.865835 &-76.865835 &  \\
50 &-76.866125 &-76.866138 &-76.866144 &-76.866146 &-76.866146 & \\
75 &-76.866449 &-76.866461 &-76.866467 &-76.866469 &-76.866469 & \\ 
100&-76.866566 &-76.866578 &-76.866584 &-76.866586 &-76.866586 & \\ 
200&-76.866682 &-76.866694 &-76.866700 &-76.866702 &-76.866702 & \\
300&-76.866703 &-76.866715 &-76.866722 &-76.866724 &-76.866724 & \\
400&-76.866711 &-76.866723 &-76.866729 &-76.866731 &-76.866731 & \\ 
500&-76.866714 &-76.866727 &-76.866733 &-76.866735 &-76.866735 & -76.866686 \\
\hline
   \end{tabular}                      
   \end{center}  
  \label{tab-entpa2}    
$^a$ See section Sec.\protect\ref{sec-tpa} for explanation.
\end{table}  
\begin{table}  
 \protect\caption{A summary of our HF results on t-PA with the STO-3G 
basis set and its comparison with
the results of other authors.  Our results are results of calculations
performed with $N=7$ and $M=500$. The bond lengths are expressed in the
units of $\AA$, the bond angles are in degrees, the total energy per C$_2$H$_2$
unit ($E_{\mbox{total}}$) is in Hartrees. The bottom four entries in this
table are results of Teramae's calculations performed using
different cutoff scheme for the Coulomb series and have been taken from
table 12 of Teramae's paper~\protect\cite{lihtpa-teramae}.}
 \protect\begin{center}  
  \begin{tabular}{ccccccc} \hline
         & $r_1$ & $r_2$ & $\Delta r$ & $R_{\mbox{CH}}$ & $\alpha$ & 
    $E_{\mbox{total}}$\\ 
Author   &       &       &            &                &        & \\ \hline
This work$^a$ &1.489 & 1.326 & 0.163      & 1.09$^b$       & 124.1  &-75.947858\\
Dovesi$^c$&1.486 & 1.329 & 0.157      & 1.09$^b$       & 124.4  &-75.946061\\
Karpfen et al.$^d$
          &1.477 & 1.327 & 0.15       & 1.09$^b$       & 124.2  &-75.948 \\
Suhai$^e$ &1.471 & 1.328 & 0.143      & 1.08           & 124.0  &-75.947283\\
Teramae$^f$
          &1.477 & 1.326 & 0.151      & 1.08           & 124.0  &-75.947935\\
Teramae$^g$
          &1.477 & 1.326 & 0.151      & 1.08           & 124.1  &-75.948581\\
Teramae$^h$
          &1.488 & 1.324 & 0.164      & 1.09           & 125.0  &-75.926695\\
Teramae$^i$
          &1.475 & 1.326 & 0.149      & 1.08           & 123.9  &-75.952922\\
\hline
   \end{tabular}                      
   \end{center}  
$^a$ Using a lobe representation of the STO-3G basis set.\\
$^b$ Held fixed at the experimental geometry~\cite{tpa-exp}. \\
$^c$ Ref.~\cite{hf-dovesi} \\
$^d$ Ref.~\cite{tpa-karp}\\
$^e$ Ref.~\cite{hf-suhai2} \\
$^f$ Obtained using the so-called Namur cutoff of the Coulomb series
     proposed by the Namur group~\cite{lih-andre}. \\
$^g$ Obtained using the cell-wise cutoff scheme for the Coulomb series
     proposed by Karpfen~\cite{cellw}. \\
$^h$ Obtained using the symmetric cutoff scheme for the Coulomb series
proposed by Kertesz et al.~\cite{symm}.\\
$^i$ Obtained using the modified symmetric cutoff scheme for the Coulomb
series proposed by Teramae himself~\cite{hf-teramae2}.
  \label{tab-sto3g}    
\end{table}  
\begin{table}  
 \protect\caption{A summary of our HF results on t-PA with the 6-31G** 
basis set and its comparison with
the corresponding calculations performed by us with the CRYSTAL program
and the results of other authors. To optimize the geometry with our
code, we performed a 
series of calculations with varying geometry 
parameters with $N=3$ and $M=75$. Experimental values are also listed for
comparison. The lengths are expressed in the
units of $\AA$, the bond angles are in degrees, the total energy per C$_2$H$_2$
unit ($E_{\mbox{total}}$) is in Hartrees while the cohesive energy per CH unit 
($E_{\mbox{coh}}$) is in eV.}
 \protect\begin{center}  
  \begin{tabular}{ccccccc} \hline
         & $r_1$ & $r_2$ & $\Delta r$ &  $\alpha$ & $E_{\mbox{total}}$ & 
         $E_{\mbox{coh}}$ \\
This work$^{a,b}$ &1.457 & 1.336 & 0.121  & 124.2     &-76.8881 & 7.32 \\
CRYSTAL$^b$       &1.464 & 1.333 & 0.131  & 124.2     &-76.8881 & 7.32 \\
Yu et al.$^{b,c}$ &1.458 & 1.335 & 0.123  & 124.1     &-76.8956 & 7.24 \\
Suhai$^d$         &1.456 & 1.339 & 0.117  & 123.9     &-76.9025 & 7.26$^e$ \\
Exp.$^f$          &1.45  & 1.36  & 0.09   & 121.7     & ---     & ---  \\ 
Exp.$^g$          &1.44  & 1.36  & 0.08   & ---       & ---     & ---  \\ 
Exp.$^h$          &1.45$\pm$0.01
                         & 1.38$\pm$0.01  
                                 & 0.07   & ---       & ---     & ---  \\ 
\hline
   \end{tabular}                      
   \end{center}  
$^a$ Performed with the lobe representation of the 6-31G** basis set described
     in the text. \\
$^b$ C-H bond distance held fixed at the experimental value 
1.09$\AA$~\cite{tpa-exp}. \\
$^c$ Ref.~\cite{corr-myu}. Yu et al. used a basis set of ``valence double
zeta + polarization'' type. \\
$^d$ Ref.~\cite{hf-suhai2}. Suhai used an extended basis set of
``double zeta + polarization'' type.  He optimized the C-H bond distance 
also to obtain 1.08 $\AA$.  \\
$^e$ Since Suhai's paper~\cite{hf-suhai2} does not provide any data on
cohesive energies, we computed it by subtracting, from his value of 
$E_{\mbox{total}}$ quoted above, the atomic HF energies of C and H computed 
with the basis set used by him.\\
$^f$ Ref. ~\protect\cite{tpa-exp} \\ 
$^g$ Ref. ~\protect\cite{tpa-exp2} \\
$^h$ Ref. ~\protect\cite{tpa-exp3} 
  \label{tab-tpaf}    
\end{table}  
\newpage
\psfig{file=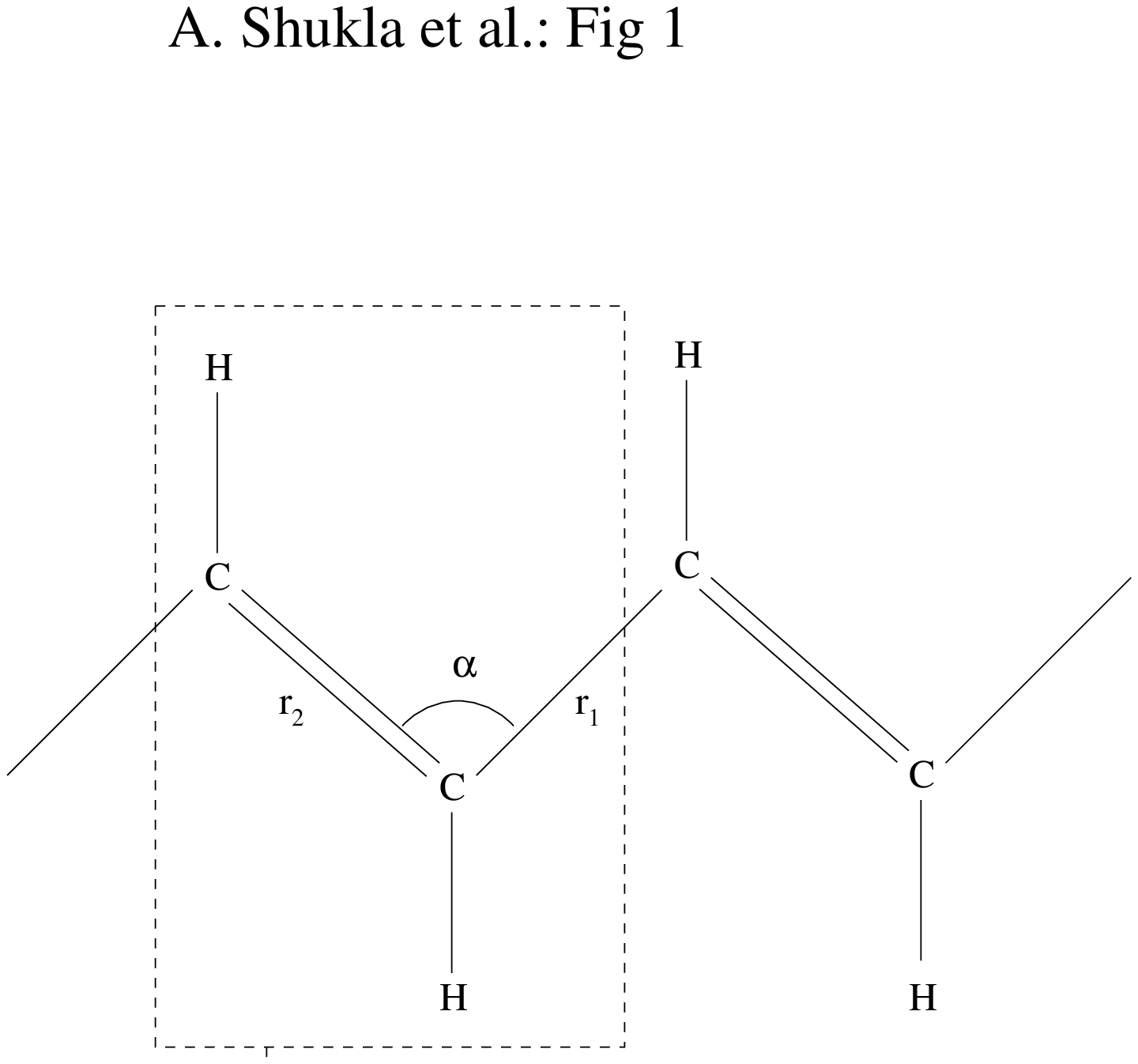,width=16cm,angle=-90}
\psfig{file=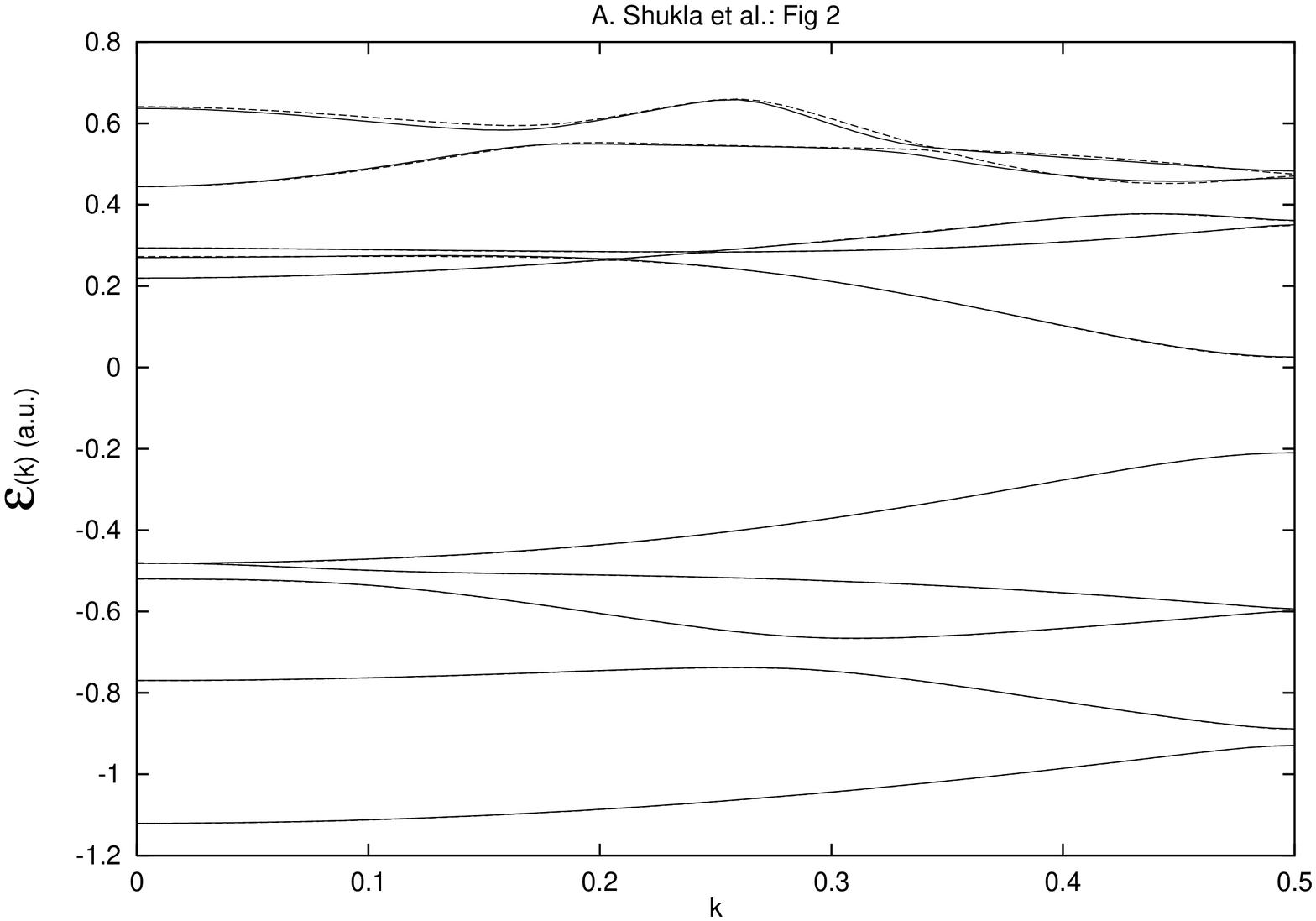,width=16cm,angle=-90}
\psfig{file=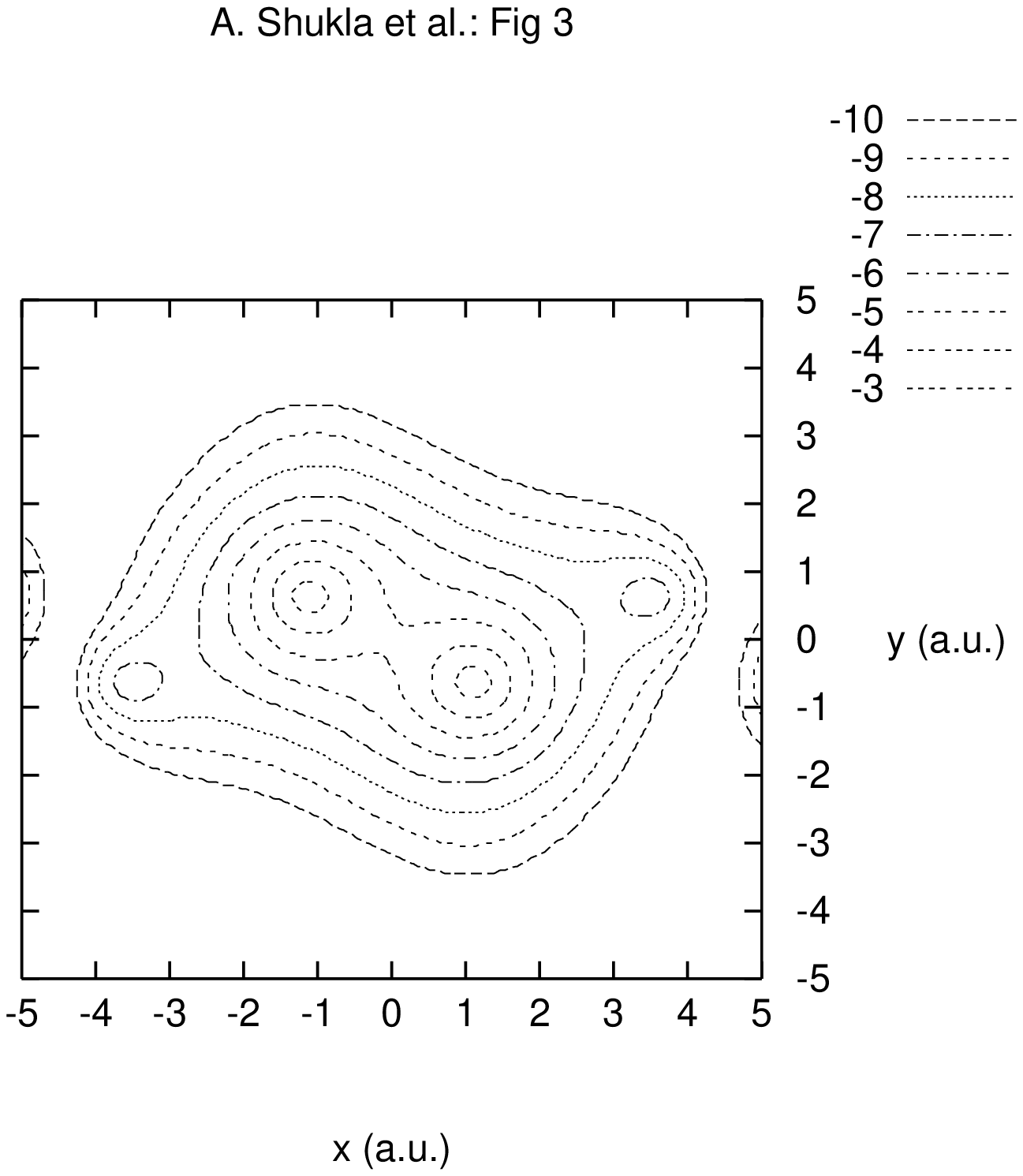,width=16cm,angle=-90}

\begin{references}
\bibitem[\dag]{email}{e-mail address: shukla@mpipks-dresden.mpg.de}
\bibitem{hf-form}{For a review of the Hartree-Fock formalism for polymers
see, e.g, G. Del Re,
J. Ladik, and G. Bizco, Phys. Rev. {\bf 155}, 997 (1967); J.M. Andr\'{e},
L. Gouverneur, and G. Leroy, Int. J. Quant. Chem. {\bf 1}, 427 (1967);
{\em ibid} 451 (1967); J.M. Andr\'{e}, J. Chem. Phys. {\bf 50}, 1536 (1969).} 
\bibitem{hf-karp}{A. Karpfen and P. Schuster, Chem. Phys. Lett. {\bf 5},
71 (1976).}
\bibitem{lih-karp}{A. Karpfen,  Theor. Chim. Acta {\bf 50}, 49 (1978).}
\bibitem{tpa-karp}{A. Karpfen and R. H\"oller, Solid State Commun. {\bf 37},
179 (1981).}
\bibitem{hf-del}{L. Piela and J. Delhalle, Int. J. Quant. Chem. {\bf 13},
 605 (1978).}
\bibitem{lih-del}{J. Delhalle, L. Piela, J.L. Br\'{e}das, and J.M. Andr\'{e},
 Phys. Rev. B {\bf 22}, 6254 (1980).}
\bibitem{tpa-kert}{M. Kert\'{e}sz, J. Koller, and A. Azman, J. Chem. Soc.
Chem. Commun. 575 (1978).}
\bibitem{hf-suhai}{S. Suhai, J. Chem. Phys. {\bf 73}, 3843 (1980).}
\bibitem{hf-suhai2}{S. Suhai, Int. J. Quant. Chem. {\bf 42}, 193 (1992).}
\bibitem{hf-dovesi}{R. Dovesi, Int. J. Quant. Chem. {\bf 26}, 197 (1984).}
\bibitem{lih-andre}{J.M. Andr\'{e}, D.P. Vercauteren, V.P. Bodart, and
J.G. Fripiat, J. Comput. Chem. {\bf 5}, 535 (1984).}
\bibitem{hf-teramae}{H. Teramae,
C. Satoko, T. Yamabe, and A. Imamura, Chem. Phys. Lett. 
{\bf 101}, 149 (1983); H. Teramae, T. Yamabe, and A. Imamura, J. Chem. Phys.
{\bf 81}, 3564 (1984).}
\bibitem{hf-teramae2}{ H. Teramae, J. Chem. Phys. {\bf 85}, 990 (1986).}
\bibitem{lihtpa-teramae}{ H. Teramae, Theor. Chim. Acta {\bf 94}, 311 (1996).}
\bibitem{corr-ladik}{S. Suhai and J. Ladik, J. Phys. C {\bf 17}, 4327 (1982).}
\bibitem{corr-suhai0}{S. Suhai, Phys. Rev. B {\bf 27}, 3506 (1983); 
Chem. Phys. Lett. {\bf 96}, 619 (1983)}
\bibitem{corr-suhai}{S. Suhai, Phys. Rev. B {\bf 29}, 4570 (1984); 
Int. J. Quant. Chem. {\bf OBS11}, 223 
(1984); J. Mol. Struct. {\bf 123}, 97 (1985); Phys. Rev. B {\bf 50}, 14791 
(1994).}
\bibitem{corr-suhai2}{S. Suhai, Phys. Rev. B {\bf 51}, 16553 (1995).}
\bibitem{corr-lieg}{C.-M. Liegener, J. Chem. Phys. {\bf 88}, 6999 (1988).}
\bibitem{corr-koenig}{G. K\"{o}nig and G. Stollhoff, Phys. Rev. Lett. 
{\bf 65}, 1239 (1990).}
\bibitem{corr-sun0}{J.Q. Sun and R.J. Bartlett, J. Chem. Phys. {\bf 104}, 8553
(1996).}
\bibitem{corr-sun}{J.Q. Sun and R.J. Bartlett, 
Phys. Rev. Lett. {\bf 77}, 3669 (1996); J. Chem. Phys. {\bf 106},
5554 (1997); J. Chem. Phys. {\bf 108}, 301 (1998); Phys. Rev. Lett. 
{\bf 80}, 349 (1998).} 
\bibitem{corr-foerner}{ Y.-J. Ye, W. F\"{o}rner, and J. Ladik, Chem. Phys.
{\bf 178}, 1 (1993); R. Knab, W. F\"{o}rner, J. \v{C}\'{\i}\v{z}ek,
 and J. Ladik, J. Mol. Struct. (Theochem.) {\bf 366}, 11 (1996), 
R. Knab, W. F\"{o}rner, and J. Ladik, J. Phys. Condens. Matter {\bf 9},
3043 (1997).}
\bibitem{corr-foerner2}{ W. F\"{o}rner, R. Knab,  J. \v{C}\'{\i}\v{z}ek,
 and J. Ladik, J. Chem. Phys. {\bf 106}, 10248 (1997).}
\bibitem{corr-myu}{M. Yu, S. Kalvoda, and M. Dolg, Chem. Phys. {\bf 224},
 121 (1997).}
\bibitem{poly-wan}{K. Fink and V. Staemmler, J. Chem. Phys. {\bf 103},
2603 (1995).}
\bibitem{shukla1}{A. Shukla, M. Dolg, H.Stoll and P. Fulde, Chem. Phys. Lett.
{\bf 262}, 213 (1996).}
\bibitem{shukla2}{A. Shukla, M. Dolg, P. Fulde, and H.Stoll, Phys. Rev.
B {\bf 57}, 1471 (1998).}
\bibitem{albrecht}{M. Albrecht, A. Shukla, M. Dolg, P. Fulde, and H.Stoll,
Chem. Phys. Lett. in press (1998).}
\bibitem{shukla3}{A. Shukla, M. Dolg, P. Fulde, and H.Stoll, J. Chem. Phys.
(in press, 1998).}
\bibitem{wannier}{Computer program  WANNIER, A.Shukla, M. Dolg, H. Stoll
and P. Fulde (unpublished).}
\bibitem{hueckel}{E. H\"uckel, Z. Phys. {\bf 70}, 204 (1931);
{\em ibid} {\bf 76}, 628 (1932).}
\bibitem{hubbard}{J. Hubbard, Proc. Roy. Soc. {\bf A276}, 238 (1963);
{\bf A277}, 237 (1964); {\bf A281}, 401 (1964).}
\bibitem{ppp}{R. Pariser and R.G. Parr, J. Chem. Phys. {\bf 21}, 767 
(1953); J.A. Pople, Trans. Farad. Soc. {\bf 49}, 1375 (1953).} 
\bibitem{ewald}{P.P. Ewald, Ann. Phys. (Leipzig) {\bf 64},  253 (1921).}
\bibitem{ew-stoll}{H. Stoll, Ph.D. thesis, Universit\"{a}t Stuttgart (1974).}
\bibitem{crystalprog}{R. Dovesi, C. Pisani, C. Roetti, M. Causa
 and V.R. Saunders, CRYSTAL88, Quantum Chemistry Program Exchange, 
Program No. 577 (Indiana University, Bloomington, IN 1989); R. Dovesi, 
V.R. Saunders and C. Roetti, CRYSTAL92 User
Document, University of Torino, Torino, and SERC  Daresbury Laboratory,
Daresbury,  UK, (1992). }
\bibitem{lih-dovesi}{R. Dovesi, C. Ermondi, E. Ferrero, C. Pisani, and
C. Roetti, Phys. Rev. B {\bf 29},  3591 (1984).}
\bibitem{huz-li}{S. Huzinaga (unpublished).}
\bibitem{huz-h}{S. Huzinaga, J. Chem. Phys. {\bf 42}, 1293 (1965).}
\bibitem{peierls}{R. Peierls, {\em Quantum Theory of Solids} (Clarendon,
Oxford, 1955), p.108.}
\bibitem{sto-3g}{W.J. Hehre, R.F. Stewart, and J.A. Pople, J. Chem.
Phys. {\bf 51}, 2657 (1969}
\bibitem{tpa-exp}{H. Kahlert, O. Leitner, G. Leising, Synthetic
Metals {\bf 17}, 467 (1987).}
\bibitem{tpa-band}{C.R. Fincher, Jr., M. Ozaki, M. Tanaka, D. Peebles,
L. Lauchlan, A.J. Heeger, and A.G. MacDiarmid, Phys. Rev. B. {\bf 20},
1589 (1979).}
\bibitem{tpa-exp2}{C.S. Yannoni and T.C. Clarke, Phys. Rev. Lett. {\bf 51},
1191 (1983).}
\bibitem{tpa-exp3}{M.J. Duijvestijn, A. Manenshijn, J. Schmidt, and R.A. Wind,
J. Mag. Reson. {\bf 64}, 451 (1985).}
\bibitem{cellw}{A. Karpfen, Int. J. Quant. Chem. {\bf 19}, 1207 (1981).}
\bibitem{symm}{M. Kertesz, J. Koller, and A. Azman, Theor. Chim. Acta
{\bf 41}, 89 (1976); M. Kertesz, Acta Phys. Hung. {\bf 41}, 107 (1976).}
\bibitem{increment}{H. Stoll, Phys. Rev. B {\bf 46}, 6700 (1992); Chem. Phys. 
Letters {\bf 191}, 548 (1992). }
\bibitem{su}{W.P. Su, J.R. Schrieffer, and A.J. Heeger, Phys. Rev. Lett.
{\bf 42}, 1698 (1979); {\em ibid}, Phys. Rev. B {\bf 22}, 2099 (1980).}
\bibitem{dixit}{See for example, S.N. Dixit and S. Mazumdar, Phys. Rev.
Lett. {\bf 51}, 292 (1983); {\em ibid}, Phys. Rev. B {\bf 29}, 1824 (1984).}
%
\end{references}
\end{document}